\documentclass[useAMS,usenatbib,usegraphicx]{mn2e}

\newcommand{\be}{\begin{equation}}
\newcommand{\ee}{\end{equation}}
\newcommand {\apgt} {\ {\raise-.5ex\hbox{$\buildrel>\over\sim$}}\ }
\newcommand {\aplt} {\ {\raise-.5ex\hbox{$\buildrel<\over\sim$}}\ }

\title[Vetting $Kepler$ Planet Candidates]{Constraining the False Positive Rate for $Kepler$ Planet Candidates with Multi-Color Photometry from the GTC}

\author[K. D. Col\'on et al.]{Knicole D. Col\'on$^{1}$\thanks{E-mail: knicole@astro.ufl.edu}\thanks{NSF Graduate Research Fellow}, Eric B.\ Ford$^{1}$, Robert C. Morehead$^{1}$\footnotemark[2]\\ 
$^{1}$Department of Astronomy, University of Florida, Gainesville, FL 32611, USA}

\begin{document}

\date{Accepted ... Received ...; in original form ...}

\pagerange{\pageref{firstpage}--\pageref{lastpage}} \pubyear{2012}

\maketitle

\label{firstpage}

\begin{abstract} 
Using the OSIRIS instrument installed on the 10.4-m Gran Telescopio Canarias (GTC) we acquired multi-color transit photometry of four small ($R_{p} \aplt 5$ $R_{\oplus}$) short-period ($P \aplt 6$ days) planet candidates recently identified by the $Kepler$ space mission.  These observations are part of a program to constrain the false positive rate for small, short-period $Kepler$ planet candidates.  Since planetary transits should be largely achromatic when observed at different wavelengths (excluding the small color changes due to stellar limb darkening), we use the observed transit color to identify candidates as either false positives (e.g., a blend with a stellar eclipsing binary either in the background/foreground or bound to the target star) or validated planets.  Our results include the identification of KOI 225.01 and KOI 1187.01 as false positives and the tentative validation of KOI 420.01 and KOI 526.01 as planets.  The probability of identifying two false positives out of a sample of four targets is less than 1\%, assuming an overall false positive rate for $Kepler$ planet candidates of 10\% \citep[as estimated by][]{morton11}.  Therefore, these results suggest a higher false positive rate for the small, short-period $Kepler$ planet candidates than has been theoretically predicted by other studies which consider the $Kepler$ planet candidate sample as a whole.  Furthermore, our results are consistent with a recent Doppler study of short-period giant $Kepler$ planet candidates \citep{santerne12}.  We also investigate how the false positive rate for our sample varies with different planetary and stellar properties.  Our results suggest that the false positive rate varies significantly with orbital period and is largest at the shortest orbital periods ($P < 3$ days), where there is a corresponding rise in the number of detached eclipsing binary stars (i.e., systems that can easily mimic planetary transits) that have been discovered by $Kepler$.  However, we do not find significant correlations between the false positive rate and other planetary or stellar properties.  Our sample size is not yet large enough to determine if orbital period plays the largest role in determining the false positive rate, but we discuss plans for future observations of additional $Kepler$ candidates and compare our program focusing on relatively faint $Kepler$ targets from the GTC with follow-up of $Kepler$ targets that has been done with warm-$Spitzer$.
\end{abstract}

\section{Introduction} 
\label{intro}

The $Kepler$ space mission has discovered, to date, 61 transiting planets as well as over 2,000 planet candidates and 2,000 eclipsing binary stars \citep{batalha12,prsa11,slawson11}.\footnote{Up to date catalogs can be found at http://kepler.nasa.gov}  With such a vast number of planet candidates, it can be difficult to decide which to focus follow-up efforts on.  Recent studies have tried to address this issue by estimating the false positive rate for the $Kepler$ sample.  Based on the list of 1,235 $Kepler$ planet candidates published by \citet{borucki11b}, it has been predicted that as many as $\sim$95\% of these candidates are true planets \citep{morton11}.  However, previous studies have not taken into account how different subsets of $Kepler$ targets may have different false positive rates.  For example, there is a rapid rise in the number of detached eclipsing binary stars that have been discovered by $Kepler$ at orbital periods of less than $\sim$ 3 days, and such systems can mimic planetary transits \citep{prsa11, slawson11}.  This suggests that there may be corresponding changes in the false positive rate with orbital period.  Because the probability of observing a transit event increases as the orbital period of the planet decreases, many of $Kepler$'s planet candidates have short periods.  Thus it is necessary to be cautious when estimating false positive rates over the whole $Kepler$ sample.  While observational studies with warm-$Spitzer$ support predictions of low false positive rates over the entire $Kepler$ sample \citep{desert12}, biases in target selection can affect observationally-constrained false positive rates (see \S\ref{discuss} for further discussion).  We also note that a recent imaging study has found that nearly 42\% of their sample of 98 $Kepler$ planet candidate hosts has a visual or bound companion within 6 arcseconds of the target star (Lillo-Box et al., in preparation).  Such studies emphasize the need for follow-up imaging to exclude blend scenarios imitating planet candidates or contaminated transit depths.

The false positive scenarios we consider in this paper are those that result from stellar eclipsing binaries that are either in the background (or, in rare cases, foreground) or bound to the target star and are not always easy to identify with $Kepler$ due to the flux from the different stars being blended together within $Kepler$'s photometric aperture ($\sim$6 arcsec).  As discussed by, e.g., \citet{colon11}, different techniques can be used to eliminate many, but not all, blends.  Multi-color transit photometry is an efficient method for recognizing blends that cannot be spatially resolved, as measuring the transit depth in different bandpasses (i.e. the transit color) allows one to test the planet hypothesis.  This is possible since the magnitude of the transit color changes as long as the blended stars have significantly different colors.    

\citet{colon11} presented multi-color transit photometry of a $Kepler$ target, KOI 565.01, that was first announced by \citet{borucki11a} to be a super-Earth-size planet candidate but later recognized as a likely false positive due to measurements of a centroid shift away from the location of the target on the CCD during transit \citep{borucki11b}.  In \citet{colon11}, we used near-simultaneous multi-color observations acquired using the narrow-band tunable filter imaging mode on the Optical System for Imaging and low Resolution Integrated Spectroscopy (OSIRIS) installed on the 10.4-m Gran Telescopio Canarias (GTC) to confirm that KOI 565.01 is indeed a false positive, as we both resolved a stellar eclipsing binary $\sim$15 arcsec from the target and measured a color change in the ``unresolved'' target+eclipsing binary system.  Thus, \citet{colon11} demonstrated the capability of the GTC/OSIRIS for efficient vetting of planet candidates via its capabilities for near-simultaneous multi-color photometry within a single transit event.  

In this paper, we present observations of four $Kepler$ planet candidates specifically selected to have small radii and short orbital periods ($R_{p} \aplt 5$ $R_{\oplus}$ and $P \aplt 6$ days).  Measuring the false positive rate for this extreme subset of planet candidates will allow us to test if there is a correlation between the false positive rate and different planetary and stellar properties.  As in \citet{colon11}, the observations presented here were acquired with the GTC/OSIRIS.  However, these observations used broadband filters in lieu of the narrow-band tunable filters in order to collect more photons and to obtain greater wavelength coverage (and thereby probe greater color differences).  In \S\ref{targets} we discuss our target selection criteria, and we discuss the corresponding observations for our four targets in \S\ref{obs}.  In \S\ref{reduction} and \S\ref{analysis} we describe our data reduction and light curve analysis procedures, and we present results for each target in \S\ref{results}.  We include a discussion of our results and how they relate to the distribution of eclipsing binaries that have been discovered by $Kepler$ as well as previous estimates of the false positive rate for the $Kepler$ sample in \S\ref{discuss}.  In particular, we discuss theoretical estimates from \citet{morton11} and observational constraints from studies by \citet{desert12} and \citet{santerne12}.  Finally, we summarize our results and conclusions in \S\ref{conc}, and we also discuss our plans for future observations of additional $Kepler$ targets with the GTC.   

\section{Target Selection}
\label{targets}

\begin{figure}
\includegraphics[width=84mm]{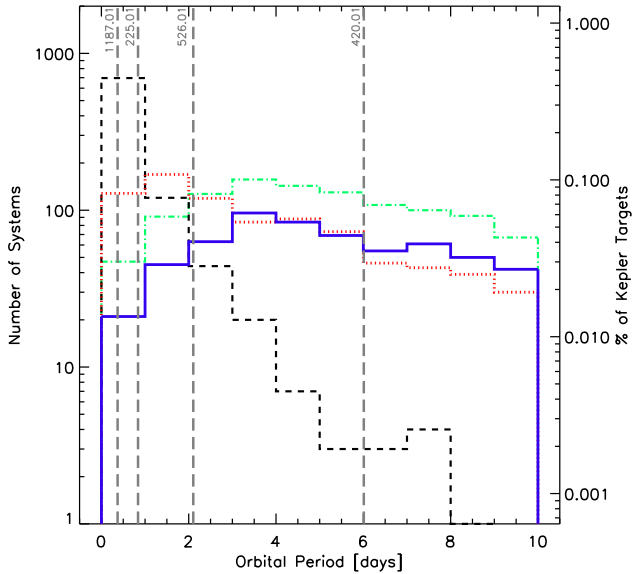}
\caption{{Histograms illustrating the number of $Kepler$ planet candidates \citep[solid blue line; based on][]{borucki11b} and detached and all ``other'' eclipsing binaries \citep[dotted red and dashed black lines, respectively; based on][]{slawson11} as a function of orbital period.  We also show a histogram (dash-dot green line) based on the recently released list of 2,321 KOIs from \citet{batalha12}.}  The righthand y-axis shows the percentage of systems relative to the total number of target stars observed by $Kepler$ in Q1.  Note that only systems with orbital periods of up to 10 days are shown.  While there is a substantial number of ``other'' (primarily semi-detached and overcontact) eclipsing binaries at the shortest orbital periods (which may be less likely to be flagged as potential planets), there is still a greater-to-comparable number of short-period detached eclipsing binaries compared to short-period planet candidates, which indicates the potential for eclipsing binaries to infiltrate the short-period planet candidate sample.  The orbital periods of the KOIs we observed are indicated with vertical long-dashed gray lines and are labeled accordingly.}
\label{perhist}
\end{figure}

Recent studies \citep[e.g.,][]{liss11, liss12} have demonstrated that a majority of the planet candidates in $Kepler$'s multi-planet candidate systems should in fact be real planets.  Considering these studies, for our program we target only those candidates found in single systems as determined by \citet{borucki11b}.  From the sample of candidates in single systems, we selected targets based on the following criteria:
\begin{itemize}
\item orbital period $(P) \aplt 6$ days
\item planet radius $(R_{p}) \aplt 6$ $R_{\oplus}$
\item transit depth (at center of transit; $\delta$) $>$ 500 ppm
\item transit duration (first to fourth contact; $\tau$) $\aplt$ 2.5 h
\item $Kepler$ magnitude (Kp) $<$ 15.5
\item vetting flag $>$ 1
\end{itemize}
As discussed in the previous section, there is a significant rise in the number of detached eclipsing binary stars compared to planet candidates at short orbital periods.  This is also illustrated in Fig. \ref{perhist}, where we show histograms of the number of planet candidates, detached eclipsing binaries, and all ``other'' eclipsing binaries \citep[i.e. all binaries that are not listed as detached in][]{slawson11} as a function of orbital period.  Therefore, we focus on constraining the false positive rate only for planet candidates with short orbital periods, where the presence of eclipsing binaries is greatest.  Our decision to focus on small candidates is due to there being a dominant population of Neptune-size or smaller candidates in the $Kepler$ sample.  We set constraints on the transit depth, transit duration and $Kepler$ magnitude due to limitations of our observing technique, so as to measure the transit depth and color precisely.  Furthermore, we want to do this while maintaining a reasonable observing cadence, hence the limits on the $Kepler$ magnitude of our targets.  Finally, we exclude candidates that have vetting flags of 1, as those have been previously confirmed as planets.

Along with the above constraints on the planetary and stellar properties, we have several observational constraints due to limitations of the GTC/OSIRIS.  First, we rule out candidates that have no observable transits from the location of the GTC (which is located on La Palma at the Observatorio del Roque de los Muchachos).  Then, we exclude those that did not have multiple transits observable during bright or grey time (during the 2011 observing season).  We also exclude transit events that occur outside altitudes of $\sim$35-72 degrees, so as to avoid high airmass and vignetting that occurs at high altitudes due to limitations of the GTC dome.  

Finally, we did $not$ check the $Kepler$ light curves\footnote{Available at http://archive.stsci.edu/kepler/} for each target that fit all of the above criteria prior to acquiring observations.  However, as we discuss in \S\ref{conc}, for future observations we plan to check the $Kepler$ light curves for secondary eclipses and V-shaped eclipses, either of which could indicate $a$  $priori$ that the target is not a transiting planet but instead a stellar eclipsing binary.

\begin{figure}
\includegraphics[width=84mm]{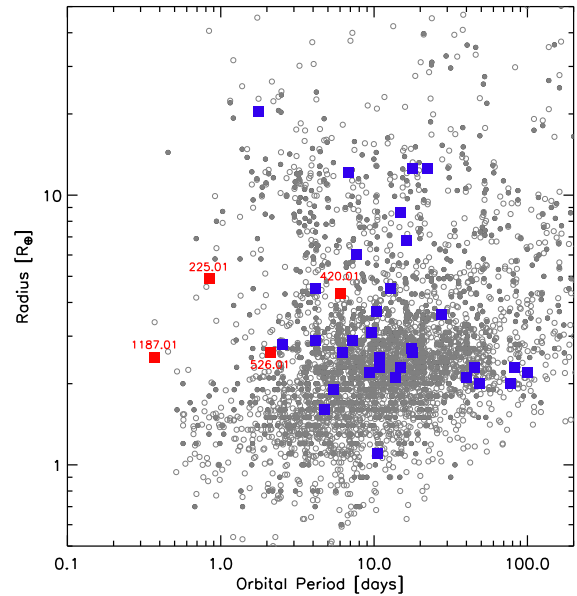}
\caption{Radius versus orbital period of $Kepler$ planet candidates.  The locations of the four KOIs we observed are indicated with filled red squares and are labeled accordingly.  The locations of the targets observed with warm-$Spitzer$ by \citet{desert12} are indicated with filled blue squares.  The filled gray circles and the open gray circles respectively represent all KOIs presented by \citet{borucki11b} and \citet{batalha12} out to orbital periods of 200 days.  Note the prominence of KOIs smaller than $\sim$ 6 $R_{\oplus}$.}
\label{radper}
\end{figure}

\begin{figure}
\includegraphics[width=84mm]{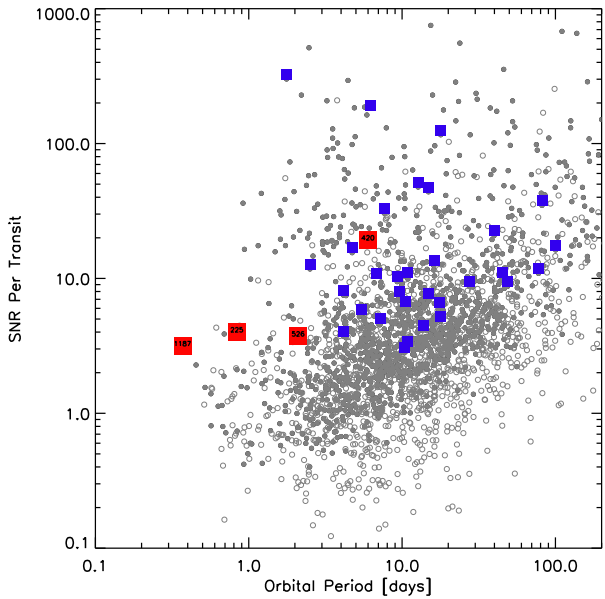}
\caption{Signal-to-noise ratio (SNR) per transit as a function of orbital period for $Kepler$ planet candidates.  The SNR per transit for each KOI was computed from the transit depth, the transit duration and either the 6-hour combined differential photometric precision (CDPP) from Q3 data (for the Borucki et al. 2011b list) or the 6-hour CDPP from Q1-Q6 data (for the Batalha et al. 2012 list).  The colors and symbols are the same as in Figure \ref{radper}.}
\label{snrper}
\end{figure}

\begin{figure}
\includegraphics[width=84mm]{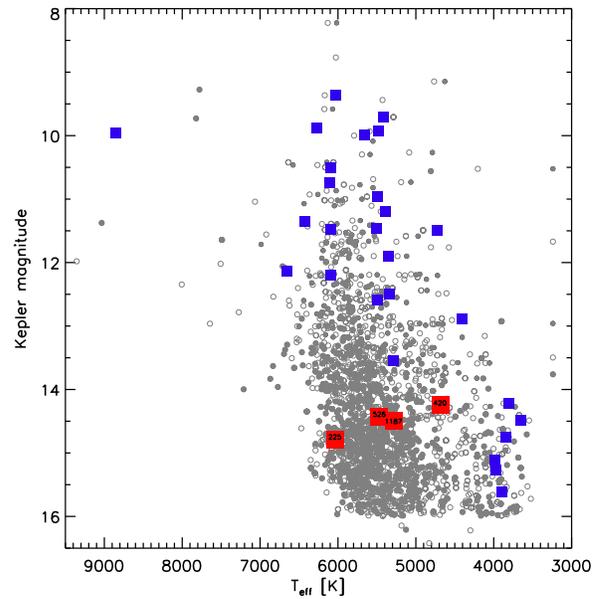}
\caption{$Kepler$ magnitude (Kp) versus effective temperature for the 997 KOI host stars published in \citet{borucki11b} and the additional 926 KOI host stars published in \citet{batalha12}.  The colors and symbols are the same as in Figs. \ref{radper} and \ref{snrper}.}
\label{kptemp}
\end{figure}

These selection criteria led us to acquire observations of four planet candidates between April and September 2011: KOI ($Kepler$ Object of Interest) 225.01, 420.01, 526.01 and 1187.01.  In Table \ref{props} we list some of the properties of each of our targets as determined by \citet{borucki11b}.  \citet{borucki11b} flagged KOI 225.01 as possibly having ellipsoidal variations.  KOI 1187.01 has the shortest period of all the planet candidates announced by \citet{borucki11b}.  In Figures \ref{radper}, \ref{snrper} and \ref{kptemp} we illustrate different properties of our targets in comparison to the sample of 1,235 KOIs and the corresponding 997 host stars from which our targets were chosen.  The sample of KOIs observed with warm-$Spitzer$ \citep{desert12} is also illustrated in these figures for comparison (we discuss the $Spitzer$ sample in more detail in \S\ref{discuss1}).  Our targets and the corresponding observations are described in detail in \S\ref{obs} below.

\begin{table*}
 \centering
 \begin{minipage}{140mm}
  \caption{KOI Properties \label{props}}
  \begin{tabular}{@{}ccccccccccc@{}}
  \hline
  \multicolumn{5}{c}{Star} &  & \multicolumn{5}{c}{Planet Candidate} \\
  \hline
KOI & KIC & Kp & T$_{eff}$ (K) & $b$ (deg) &  & $P$ (days) & $\tau$ (hr) & $\delta$ (ppm) & $R_{p}$ ($R_{\oplus}$) & $R_{p}/R_{\star}$ \\
\hline
225   & 5801571 & 14.784 & 6037 & 9.177   &   & 0.838598    & 1.2452 & 2571 & 4.9 & 0.04932 \\ 
420   & 8352537 & 14.247 & 4687 & 16.412 &   & 6.010401    & 2.2582 & 2700 & 4.3 & 0.0474 \\ 
526   & 9157634 & 14.427 & 5467 & 12.426 &   & 2.104719    & 1.7639 & 926   & 2.6 & 0.0305 \\
1187 & 3848972 & 14.489 & 5286 & 10.846 &   & 0.3705285 & 0.7778 & 1835 & 2.5 & 0.03961 \\ 
\hline
\end{tabular}

\medskip
All values are from \citet{borucki11b}.  The KIC number refers to the $Kepler$ Input Catalog number for each target.  Note that $b$ is the Galactic latitude of the KOI host star, $\tau$ is the transit duration and $\delta$ is the transit depth as measured in the $Kepler$ bandpass.

\end{minipage}
\end{table*}


\section{Observations}
\label{obs}

We acquired photometry of each target and several nearby reference stars around the predicted time of a transit event.  For each observation, we used the GTC/OSIRIS to acquire near-simultaneous multi-color photometry by alternating between two broadband ``order sorter'' filters\footnote{http://www.gtc.iac.es/en/pages/instrumentation/osiris.php} that were custom made for OSIRIS: 666$\pm$36 nm and 858$\pm$58 nm.  Both bandpasses were specifically chosen so as to minimize effects of telluric absorption and emission while also allowing for ample wavelength coverage.  In order to decrease dead time, all observations used 1$\times$1 binning, a fast pixel readout rate of 500 kHz and a single window located on one CCD chip.  The size of the window varied for each observation, but each window was large enough to contain the target and several reference stars.  We note that all observations except those for KOI 1187 were conducted in queue (service) mode.  In the following sections, we describe specific details regarding each target and its respective observations.   

\subsection{KOI 225.01}
\label{koi225obs}

We observed the transit event of KOI 225.01 on 2011 April 13 under clear conditions and during dark time, with observations beginning at 03:10 UT and ending at 06:12 UT, during which time the airmass ranged from $\sim$ 1.66 to 1.07.  The exposure time was 50 s (for both bandpasses), with $\sim$40 s of dead time following each exposure.  Seeing varied between 1.2 and 2.0 arcsec until about 05:00 UT, at which point the seeing improved to $<$ 1 arcsec and a slight defocus was introduced in order to avoid saturation.  A diffuse, dark band was present in all the images (a result of a very bright star located just outside the CCD), so the target was positioned on the CCD such that it was outside this region and also avoided bad pixels in the center of the CCD chip.  During the observations, the target's centroid coordinates shifted by $<$ 9 pixels in the $x$-direction and $\sim$ 2 pixels in the $y$-direction.  Several images were lost due to technical issues, and a few of the images towards the end of the observations were discarded due to the beginning of twilight.  

\subsection{KOI 420.01}
\label{koi420obs}

Observations of the 2011 September 13 transit of KOI 420.01 took place from 21:48 UT (2011 September 13)  to 01:30 UT (2011 September 14).  Observations occurred during bright time and under photometric conditions, and the airmass ranged from 1.07 to 1.84.  The exposure time was set to 40 s, with a corresponding 20 s of dead time following each exposure.  The seeing roughly ranged from 1.1-1.8 arcsec throughout the observations, though at the beginning of the observations the seeing conditions changed drastically enough from one image to another that some of the reference stars were saturated.  We explicitly exclude any saturated reference stars from our analysis (see \S\ref{analysis} for further details).  The seeing stabilized and improved after about 23:00 UT, enough so that a slight defocus was implemented to avoid saturation.  The target's centroid coordinates shifted by $\aplt$ 4 pixels in either direction during the observations.

\subsection{KOI 526.01}
\label{koi526obs}

The 2011 September 11 transit of KOI 526.01 was observed under photometric conditions and during bright time.  Observations began at 21:37 UT on 2011 September 11 and ended at 00:00 UT on 2011 September 12.  The airmass ranged from 1.05 to 1.22 and the seeing was stable between 0.6 and 0.8 arcsec.  A slight defocus was implemented, yielding a defocused FWHM of 0.8-1.0 arcsec.  An exposure time of 10 s was used, with 20 s of dead time between exposures.  As in observations of KOI 225.01 (\S\ref{koi225obs}), a dark band caused by a bright star outside the CCD window was present in all images, so we placed the target star appropriately far away from the band so that the photometry would not be affected.  The centroid coordinates of the target shifted by less than 4 and 2 pixels in the $x$- and $y$-directions, respectively.  Twice during the observations {the primary mirror segments lost alignment and produced distorted images}: first around 22:30 UT and again at 00:00 UT.  We discarded a few images from the first instance, but the issue with the mirror could not be readily fixed during the second instance, and the observations were forced to end early, around the time of the transit egress.  

\subsection{KOI 1187.01}
\label{koi1187obs}

We observed the 2011 June 12 transit event of KOI 1187.01, with observations taking place from 22:31 UT (2011 June 11) to 02:01 UT (2011 June 12).  Observations took place during bright time and under clear conditions, and the airmass ranged from 1.90 to 1.05.  Seeing was excellent and ranged from 0.6 to 0.9 arcsec during the night.  Due to the excellent seeing conditions, the exposure time was set to 5 s.  There were 24 s of dead time following each exposure.  During the observations, the target's centroid coordinates shifted by $\aplt$ 4 pixels in either direction.     

\section{Data Reduction}
\label{reduction}

\begin{figure}
\includegraphics[width=84mm]{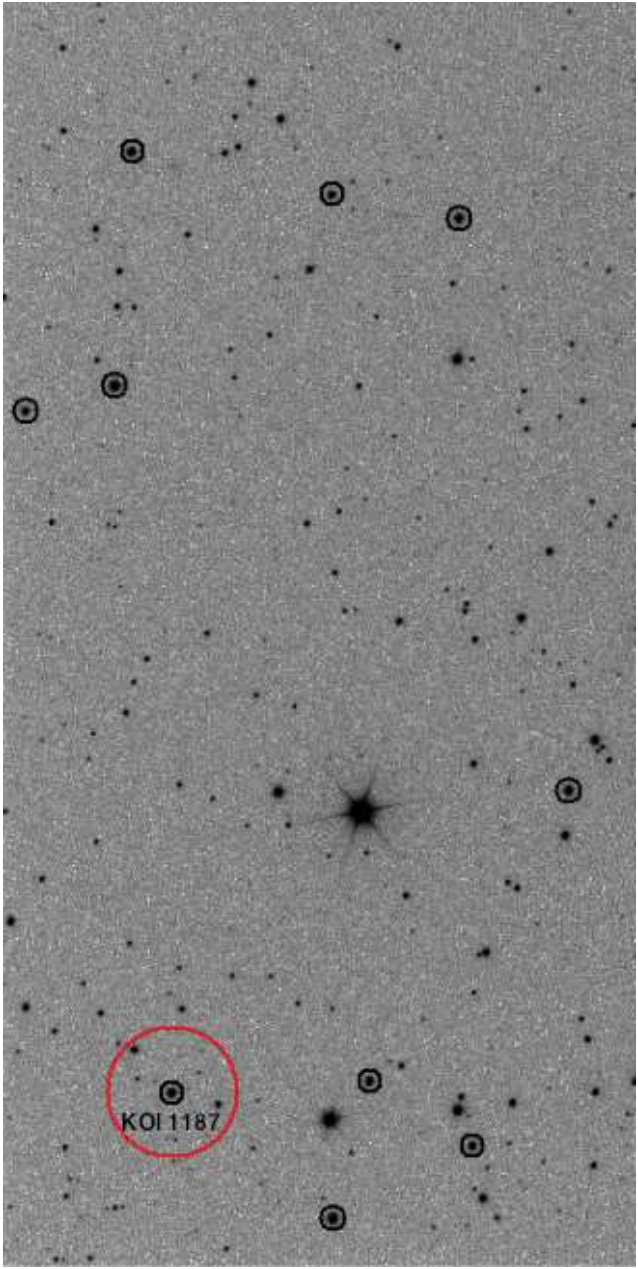}
\caption{Image of the field of view around KOI 1187 as acquired using the GTC/OSIRIS and the 666/36 nm order sorter filter.  The target is labeled accordingly and is contained within a black circle equivalent to twice the aperture used in the reduction process.  The unlabeled stars contained in black circles are the nine reference stars used in our analysis.  The larger red circle around KOI 1187 has a radius of $\sim$ 20 arcsec, and the six brightest stars located within this circle (excluding the target) are the stars that we investigated as potential sources of the transit signal.}
\label{fov}
\end{figure}

We used standard {\sc iraf} procedures for bias subtraction and flat-field correction (using dome flats taken for each bandpass) for each target.  {Due to nonuniform illumination by the lamp, we added an illumination correction to the final flat-field image for each target (except KOI 225).  The illumination correction was performed using the {\sc iraf} task $mkillumflat$ within the $noao.imred.ccdred$ package, which removes the large scale illumination pattern from the flat field by smoothing the flat field image.  For KOI 225}, the final flat-field for the 666 nm filter showed a strong gradient after performing an illumination correction, so we chose to use the flat-field image as it was.  To be consistent, we did not perform an illumination correction on the 858 nm final flat-field for KOI 225 either.  {We used the IDL Astronomy User's Library\footnote{http://idlastro.gsfc.nasa.gov/} implementation of {\sc daophot} \citep{stetson}} to perform aperture photometry on each target and several nearby reference stars as well as on other potential sources of the transit signal (i.e. stars within $\sim$ 20 arcsec of each target).  For each target, we tested different apertures and chose a final aperture based on that which resulted in the smallest scatter in the baseline (out-of-transit) flux ratios (i.e., the target star flux divided by the ensemble reference star flux).  For KOI 225, 420, 526 and 1187 our final aperture was 25, 23, 15 and 14 pixels (equivalent to approximately 3.2, 2.9, 1.9 and 1.8 arcsec).  Sky background subtraction took place during the aperture photometry process, with annuli chosen to be far enough away from each star that the flux from a given star would not be included within the sky annulus.  While these procedures were performed for each bandpass separately, we considered the results for each bandpass and used the same aperture and sky annulus for each data set for a given target.  Note that in many cases a smaller aperture and sky annulus were used for the stars within $\sim$ 20 arcsec of a given target due to the small separations between some of the stars and the potential for blending.  For reference, we present a reduced science image showing the field of view around KOI 1187 in Figure \ref{fov}.  Once aperture photometry and sky subtraction were completed, we proceeded with light curve analysis as discussed in \S\ref{analysis} below.    

\section{Light Curve Analysis}
\label{analysis}

For each target, we computed light curves for each bandpass by dividing the total flux measured from the target by the total weighted flux of an ensemble of reference stars, where the flux from each reference star was weighted accordingly after discarding outlying flux values (resulting from either technical issues or variable sky conditions).  During the analysis, several reference stars were found to be obviously variable or saturated, so we excluded those stars from the analysis.  This resulted in us using 10, 4, 2 and 9 reference stars in the field near KOI 225, 420, 526 and 1187 to compute the reference ensemble flux.  Next, each light curve for each target was normalized to the mean baseline flux ratio as measured in each bandpass.  The light curves were then corrected for changes in airmass as well as for drifts in the centroid coordinates of the target on the CCD and the sharpness of the target's profile [(2.35/FWHM)$^{2}$].  The latter corrections were done via external parameter decorrelation \citep[see, e.g.,][]{bakos07,bakos10}.  We note that due to the lack of data for KOI 526 during the transit egress and post-transit, our attempts to correct those light curves produced skewed results.  Therefore, all further analysis for KOI 526 is based on the uncorrected light curves.  

The corresponding observation times for each light curve for each target were computed from the UTC timestamps given in the image headers.  We converted the UTC time at mid-exposure to Barycentric Julian Dates in Barycentric Dynamical Time (BJD\_TDB) via an online calculator described in \citet{eastman10} in order to match the time coordinate system that the ephemerides from \citet{borucki11b} are given in.\footnote{The calculator is available at http://astroutils.astronomy.ohio-state.edu/time/utc2bjd.html}  

The photometric uncertainties were computed from the photon noise of the target star and the reference star ensemble, the noise in the sky background around the target and each reference used in the ensemble, and scintillation noise.  The resulting median photometric uncertainties for each light curve for each target are given in Table \ref{errs}.  In each case, the photon noise of the target is the primary source of error.  Finally, in order to investigate the presence of red noise in our data, we computed the standard deviation of the flux ratios for different bin sizes and found that all our data follow the trend expected for white (Gaussian) noise.  Despite this result, we also investigate potential residual systematics via a ``prayer-bead'' analysis, which we discuss below.

\begin{table}
 \centering
  \caption{Photometric Precisions \label{errs} }
  \begin{tabular}{@{}ccc@{}}
  \hline
  KOI & $\sigma_{666nm}$ (ppm) & $\sigma_{858nm}$ (ppm) \\
  \hline
  225 & 476 & 540 \\
  420 & 501 & 428 \\
  526 & 1223 & 1100 \\
  1187 & 1447 & 1295 \\
  \hline
  \end{tabular}

\end{table}

After correcting the light curves for each target (except for KOI 526 as discussed above), we proceeded to fit synthetic models to each light curve, following the approach taken by \citet{colon10}.  We assumed each observed transit event was a planetary transit, and we used the planetary transit light curve models from \citet{mandel} to fit limb-darkened models to our data.  Specifically, for each target, we fit for the following parameters:
\begin{itemize}
\item time of mid-transit ($t_0$)
\item transit duration (first to fourth contact; $\tau$)
\item impact parameter ($a \cos i/R_{\star}$)
\item planet-star radius ratio ($R_p/R_{\star}$)
\item two limb darkening coefficients ($c_1$ and $c_2$)
\item baseline flux ratio
\item (linear) baseline slope
\end{itemize}
The limb darkening coefficients that we fit for are defined as $c_1 \equiv u_1 + u_2$ and $c_2 \equiv u_1 - u_2$, where $u_1$ and $u_2$ are linear and quadratic limb darkening coefficients.  For KOI 526, we note that we held the transit duration fixed to the value from \citet{borucki11b} in order to fit a model to our partial transit.

Initial guesses for the mid-transit time, transit duration, impact parameter and radius ratio were based on the values and their corresponding uncertainties as given in \citet{borucki11b}.  Values for the limb darkening coefficients were interpolated from the \citet{claret11} models for the Sloan $r^{'}$ and $z^{'}$ filters and are based on the stellar parameters given in \citet{borucki11b}.  Due to the potentially larger than estimated uncertainties in the stellar parameters \citep[see, e.g.,][]{borucki11b}, we did not allow the limb darkening coefficients to be completely free parameters in the fitting process.  Rather, we kept the coefficients fixed at self-consistent values during the fitting process, but we tested a range of fixed values for the coefficients.  Similarly, we tested a range of initial guesses for the other parameters based on the uncertainties for the planetary and stellar parameters.  Best-fitting models were identified via a Levenberg-Marquardt minimization scheme.\footnote{We specifically used $mpfitfun$, which is publicly available at http://www.physics.wisc.edu/$\sim$craigm/idl/idl.html}  

Our light curve fitting procedure for each target is as follows.  First, we fit models to each light curve individually, corrected the data against the best-fit baseline flux ratio and slope, subtracted the best-fit models and discarded any data points lying greater than 3$\sigma$ from the residuals.  Then, we used the corrected light curves and fit models to them in a joint analysis, where we forced the impact parameter, transit duration, mid-transit time, baseline flux ratio and baseline slope to be the same for both light curves.  However, we allowed different values for the radius ratio and limb darkening coefficients to be fitted to the different light curves.  The results from the joint analysis were then used to correct the individual light curves as in the first step described above.  A final joint analysis was then applied to the ``final'' corrected light curves.  During this final stage, we also performed a ``prayer-bead'' analysis as was done in \citet{colon10}.  Specifically, we performed a circular shift on the residuals for each light curve (computed after removing the best-fit model) and constructed synthetic light curves by adding the shifted residuals back to the best-fit model.  The joint analysis described above was applied to each synthetic light curve, and we use the dispersion of the best-fit parameters to calculate uncertainties on each parameter.  This accounts for any additional systematic noise sources in the data.  We present results from our light curve analysis in \S\ref{results}.

\section{Results}
\label{results}

\begin{figure}
\includegraphics[width=84mm]{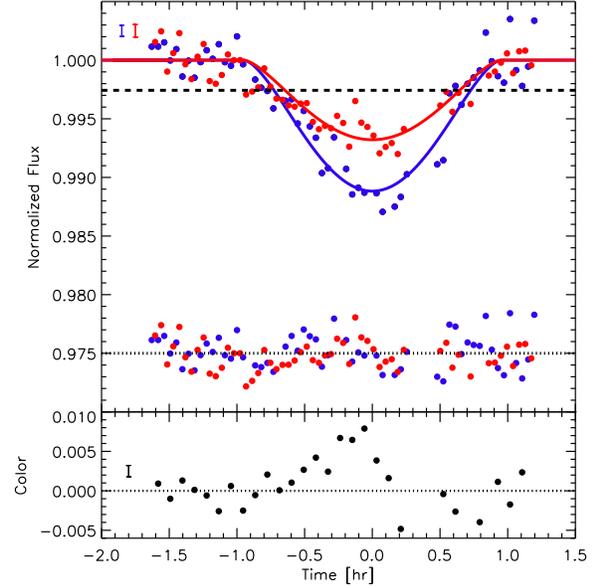}
\caption{Light curves, residuals, and color based on observations of the 2011 April 13 transit of KOI 225.01 acquired at 666 nm (in blue) and 858 nm (in red).  The circles are observations and the solid curves are the corresponding best-fit models (see text for further details).  Representative photometric error bars for each light curve are shown on the left-hand side of the top panel.  The black horizontal dashed line illustrates the depth of the transit as measured in the $Kepler$ bandpass (see Table \ref{props}).  Residuals from the fits for each light curve are also shown in the top panel, with a horizontal dotted line indicating the level at which the residuals were offset (for clarity).  The bottom panel shows the color as computed from the two light curves as well as a representative error bar on the left-hand side of the panel and a horizontal dotted line that illustrates a color of zero (for reference).}
\label{lc225}
\end{figure}

\begin{figure}
\includegraphics[width=84mm]{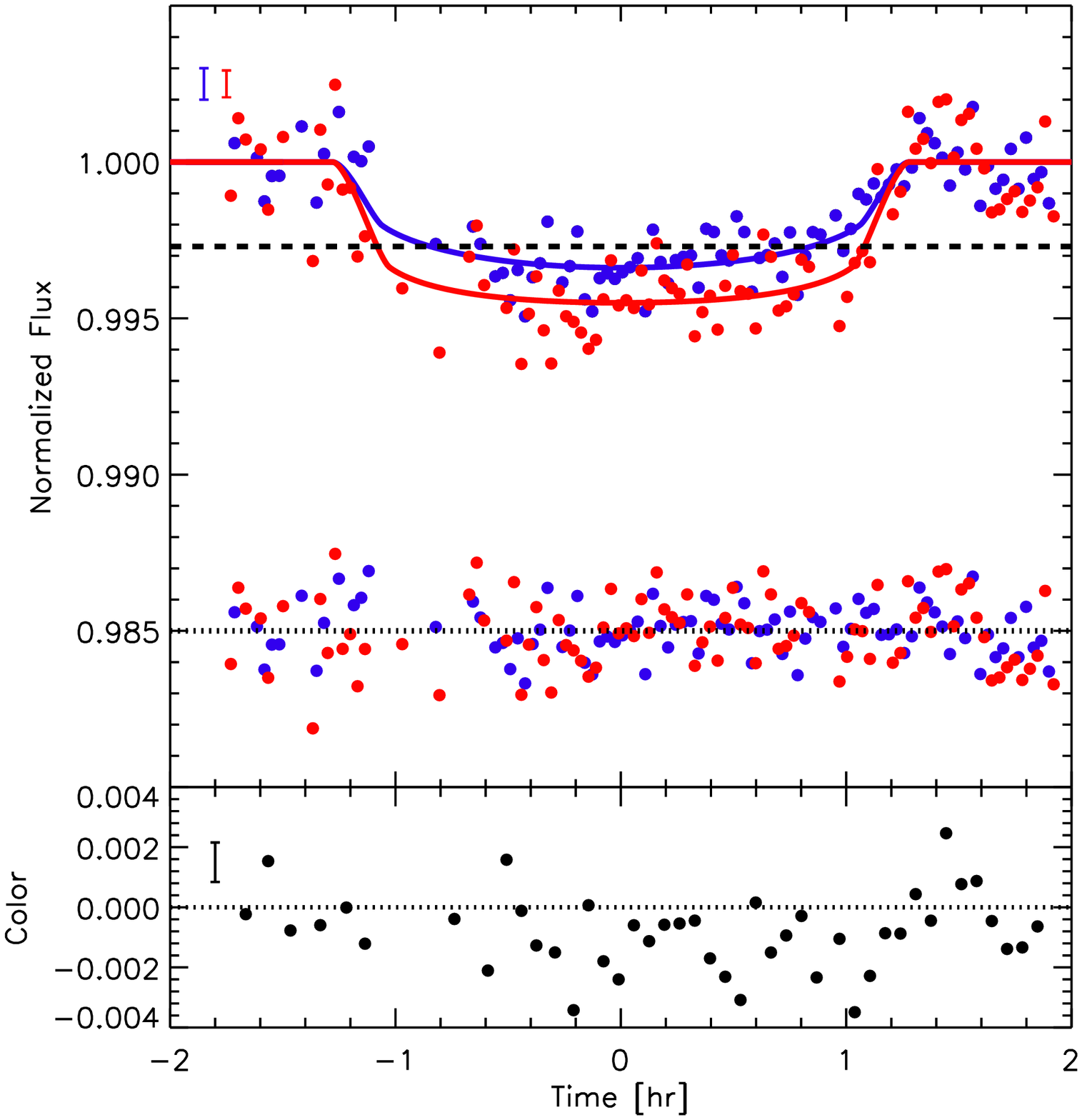}
\caption{Same as Figure \ref{lc225}, but for the transit of KOI 420.01 as observed on 2011 September 13.}
\label{lc420}
\end{figure}

\begin{figure}
\includegraphics[width=84mm]{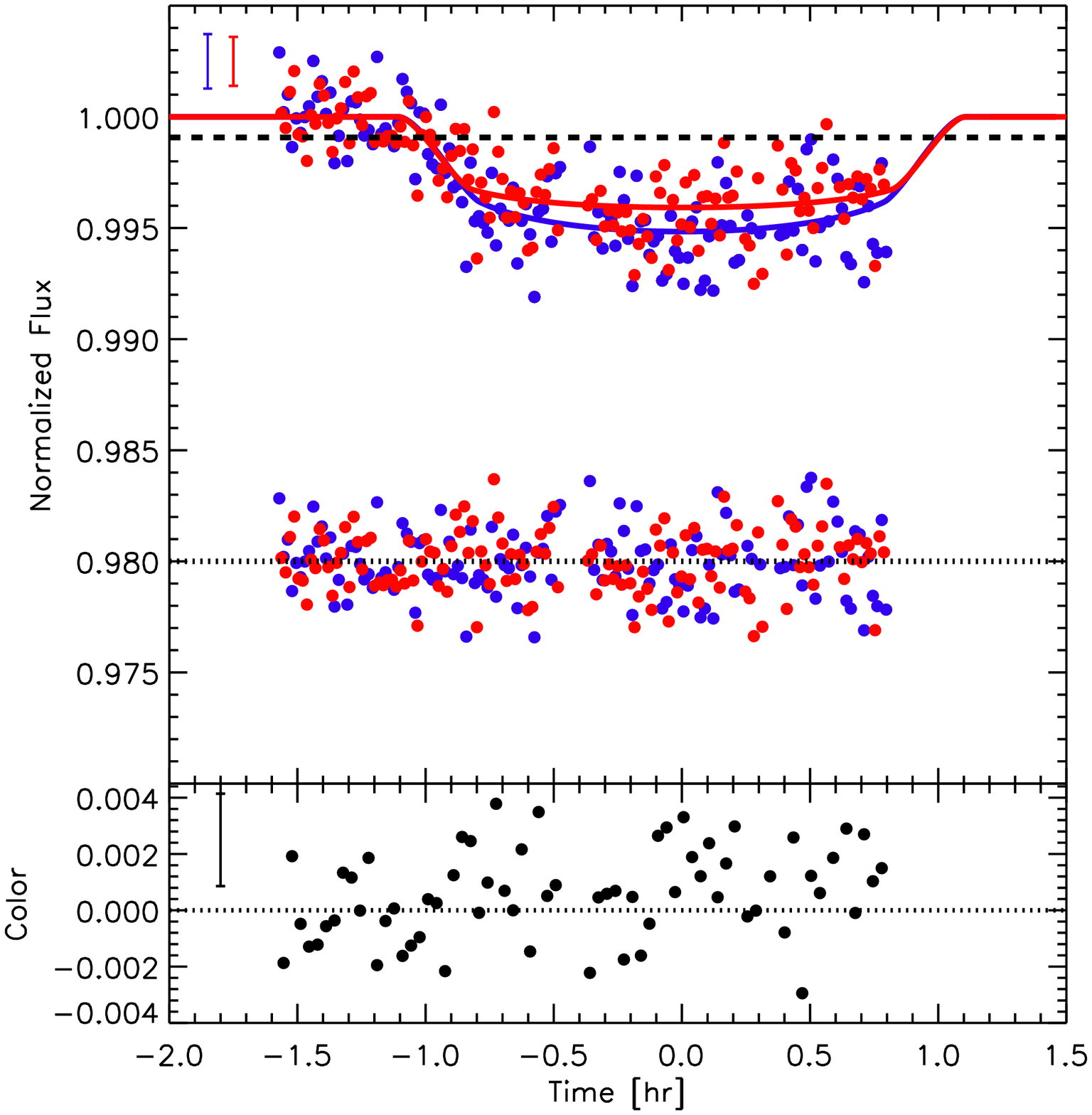}
\caption{Same as Figure \ref{lc225}, but for the transit of KOI 526.01 as observed on 2011 September 11.}
\label{lc526}
\end{figure}

\begin{figure}
\includegraphics[width=84mm]{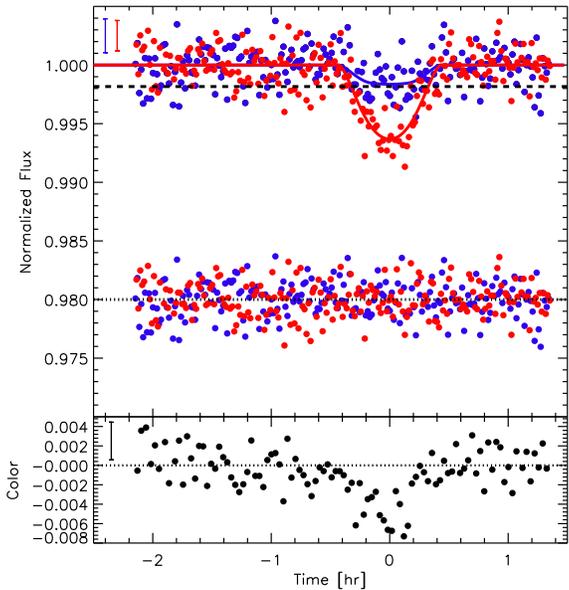}
\caption{Same as Figure \ref{lc225}, but for the transit of KOI 1187.01 as observed on 2011 June 12.}
\label{lc1187}
\end{figure}

We present the light curves and the corresponding best-fit models for each target in Figures \ref{lc225}, \ref{lc420}, \ref{lc526} and \ref{lc1187}.  While not shown here, we had also generated light curves for the potential sources of each transit signal (i.e. stars that could have been blended with the target within $Kepler$'s aperture) following a similar procedure as described above.  We generated light curves for 10, 3, 3 and 6 stars within $\sim$ 20 arcsec of KOI 225, 420, 526 and 1187.  Upon visual inspection, we found that none of these light curves showed a transit signal during the time of the transit event, indicating that the transits that we observed either occur due to an object transiting the target star or an object transiting an unresolved star that is blended with the target star.  We also present the transit color (666 nm $-$ 858 nm) of each target in the bottom panel of Figures \ref{lc225}$-$\ref{lc1187}, which was computed by taking the average of each pair of flux ratios in the 666 nm light curve and dividing by the corresponding points in the 858 nm light curve.  Therefore, in Figures \ref{lc225}$-$\ref{lc1187}, a positive color indicates a redder transit and a negative color indicates a bluer transit.

In Table \ref{fits}, we report best-fit model parameters for each target based on the model with the smallest $\chi^2$ value, but we note that the parameter uncertainties are based on the full set of best-fit values and their corresponding uncertainties as found during the ``prayer-bead'' analysis.  Specifically, for a given parameter, the upper error bar is a result of subtracting the best-fit value (from the model with the smallest $\chi^2$ value) from the maximum sum of a fitted value and its associated measurement uncertainty as determined from the full set of models computed during the ``prayer-bead'' analysis, and likewise for the lower error bar.  In general, we find that the formal 1$\sigma$ errors for the best-fit parameters are comparable to those errors computed based on the results from the ``prayer-bead'' analysis, which indicates that any residual systematics in the data (i.e. those not removed by external parameter decorrelation or airmass corrections) have a negligible effect on our results.  Furthermore, the distribution of each best-fit light curve parameter over all permutations is smaller than the uncertainties given in Table \ref{fits}, which further indicates that the results presented here are robust.  

{Finally, as suggested by the referee, we consider whether we can use astrometry from our images to provide additional constraints on the properties of the targets we observed.  A preliminary astrometric analysis does not suggest that any of the KOIs we observed are due to a blend with another star that was resolved by our observations.  Furthermore, despite the large aperture used by $Kepler$, the astrometric precisions from $Kepler$ are extremely high ($<$ 0.004 arcseconds for a single 30 min exposure; \citealp{monet10}), and to the best of our knowledge, no significant centroid shifts were measured by $Kepler$ for any of our KOIs.}

We discuss specific results for each target individually in the following sections.

{\begin{table*}
 \centering
 \begin{minipage}{180mm}
  \caption{Best-Fit Model Parameters \label{fits}}
  \begin{tabular}{@{}cccccccccc@{}}
  \hline
KOI & $t_0$ & $\tau$ & $a \cos i/R_{\star}$ & $R_{p}/R_{\star}$ & $R_{p}/R_{\star}$ & $c_1$ & $c_2$  & $c_1$ & $c_2$ \\
 & (BJD$-$2454900) & (hr) &  & (666 nm)  & (858 nm) & (666 nm) & (666 nm) & (858 nm) & (858 nm) \\
\hline
 225.01  &  764.70659$^{+0.00096}_{-0.00026}$  &  1.7466$^{+0.0214}_{-0.0871}$  &  1.0  &  0.1739$^{+0.0034}_{-0.0012}$  &  0.1318$^{+0.0026}_{-0.0015}$  &   0.6548  &   0.0152  &   0.5034  &  -0.0398  \\
420.01  &  918.49094$^{+0.00039}_{-0.00102}$  &  2.1188$^{+0.0560}_{-0.0379}$  &  0.6270$^{+0.0898}_{-0.1498}$  &  0.0552$^{+0.0021}_{-0.0021}$  &  0.0647$^{+0.0014}_{-0.0013}$  &   0.7581  &   0.4729  &   0.5837  &  -0.2011  \\
526.01  &  916.46330$^{+0.00141}_{-0.00170}$  &  1.7639 (fixed)  &  0.7456$^{+0.0771}_{-0.0771}$  &  0.0714$^{+0.0037}_{-0.0029}$  &  0.0635$^{+0.0035}_{-0.0025}$  &   0.6920  &   0.2466  &   0.5339  &   0.0869  \\
1187.01  &  824.52863$^{+0.00049}_{-0.00045}$  &  0.5066$^{+0.0390}_{-0.0285}$  &  0.9345$^{+0.0427}_{-0.0423}$  &  0.0469$^{+0.0074}_{-0.0070}$  &  0.0924$^{+0.0174}_{-0.0173}$  &   0.6895  &   0.2387  &   0.5370  &   0.0886  \\
\hline
\end{tabular}

\medskip
Note that the time of mid-transit, $t_0$, is technically given in BJD\_TDB (Barycentric Julian Date in Barycentric Dynamical Time).  See text for further details.

\end{minipage}
\end{table*}


\subsection{KOI 225.01}
\label{koi225res}

The light curves presented in Figure \ref{lc225} are prominently V-shaped, which alone suggests a possible non-planetary transit event.  Considering the shape of the light curves and the significant difference in the transit depths as measured in the GTC bandpasses, we deduce that KOI 225.01 is most likely a stellar eclipsing binary system.  Specifically, we find that the best-fit planet-star radius ratios differ at a level of $>$ 11$\sigma$, which is a clear indication that this object is not a planet.  The transit depth in the $Kepler$ bandpass is also significantly different from those measured in the GTC bandpasses.  {In particular, the transit depth is notably different between the $Kepler$ bandpass and the blue GTC bandpass (around 666 nm), even though they probe fairly similar wavelength regimes.  We primarily attribute this difference to dilution in the $Kepler$ photometry, since there is a relatively bright star near KOI 225 that appears to have slightly contaminated one of the pixels used in the $Kepler$ photometry.\footnote{As determined from target pixel files downloaded from http://archive.stsci.edu/kepler/}}  As noted in \citet{borucki11b}, this target's light curve had possible ellipsoidal variations, also suggesting that the system contains an eclipsing binary.  Furthermore, after our observations had been conducted, \citet{slawson11} listed this target in their eclipsing binary catalog.\footnote{We refer the reader to \citet{slawson11} for further details on how some KOIs were rejected as planet candidates and subsequently added to the eclipsing binary catalog.}  {Finally, \citet{ofir12} recently conducted an independent analysis of the $Kepler$ data set and identified KOI 225.01 as a false positive due to the detection of significant differences between the odd and even eclipse events.} 

Since we measure a deeper eclipse in the bluer GTC bandpass, this indicates that during the eclipse more blue light is blocked, and therefore the secondary (eclipsing) component is redder than the primary star.  In Table \ref{fits}, we provide an updated ephemeris and eclipse duration, though we note that likely due to the non-planetary transit light curve shape, the best-fit model yielded an impact parameter that reached the upper boundary limits during the fitting process.  Therefore, the resulting best-fit light curve parameters are likely somewhat skewed, but this does not change the significance of our results.  

\citet{slawson11} report a period that is twice as long as that found by \citet{borucki11b}.  If correct, then either what \citet{borucki11b} believed were transit events were actually a combination of primary and secondary eclipses  $or$ the eclipsing binary has a large enough inclination that all eclipse events are primary eclipses and the secondaries are unobservable from our line-of-sight.  If we assume that the period found by \citet{slawson11} is correct (and was determined, for instance, by measuring different depths for successive transits, which led to this KOI being rejected), then it is necessary to consider how this affects any correlation between the orbital period and the false positive rate for $Kepler$ targets.  We refer the reader to \S\ref{discuss} for further discussion.

\subsection{KOI 420.01}
\label{koi420res}

As illustrated in Figure \ref{lc420}, the transit depths as observed in the GTC bandpasses are comparable, and there is not a significant change in the measured color during transit.  The planet-star radius ratios measured from the GTC light curves are consistent within $\sim$ 2.8$\sigma$, and we did not resolve any potential sources of the transit signal within $\sim$ 20 arcsec of the target, so we deduce that this target is {\it not} a false positive and is instead a {\it validated} planet.\footnote{We use the term {\it validated} to mean that the planet candidate is most likely a planet based on our observations but it is not a {\it confirmed} planet because there is not independent evidence (e.g., Doppler or transit timing variations) for the planet model.}  The best-fit parameters from our light curve models are given in Table \ref{fits} and include an updated ephemeris, transit duration, and impact parameter.  We note that the impact parameter given in \citet{borucki11b} has an associated uncertainty of 1, and {\citet{batalha12} report an impact parameter of 0.57 with a relatively large uncertainty of 0.52}, so we provide a much stronger constraint on this parameter here.  Finally, based on our measured planet-star radius ratios and the revised stellar radius (0.69 $R_{\odot}$ compared to the value of 0.83 $R_{\odot}$ reported by \citealp{borucki11b}) given in \citet{batalha12}, we find KOI 420.01 to have a radius that is between $\sim$ 4.15 and 4.87 $R_{\oplus}$ (slightly larger than the value of 3.65 $R_{\oplus}$ found by \citealp{batalha12}).

\subsection{KOI 526.01}
\label{koi526res}

Despite only observing a partial transit with the GTC (as illustrated in Figure \ref{lc526}), we find that the transit depths clearly match each other, with planet-star radius ratios that are consistent within $\sim$ 1.2$\sigma$.  {While the $Kepler$ depth is much shallower, there appears to be no significant contamination from nearby stars in the $Kepler$ photometry (see footnote in \S\ref{koi225res}).  Given that even the transit depths in the $Kepler$ and the blue GTC bandpass are significantly different, and that dilution appears to not be the source, we believe this object warrants further investigation beyond the scope of this paper.  For instance, it is possible that stellar variability in the target star or a blended star impacted both the GTC and $Kepler$'s measured transit depths.\footnote{$Kepler$ light curves for this target show baseline variability at a level of $\sim$1\%.}  Regardless, since the depths between the GTC bandpasses are consistent, this target still passes our validation test.}  As in the case for KOI 420.01, we found no stars within $\sim$ 20 arcsec of the target that showed a transit signal at the expected time of the transit, so we also {\it validate} KOI 526.01 as a planet.  Table \ref{fits} includes an updated ephemeris and impact parameter (recall that the transit duration was held fixed due to fitting only a partial transit).  Again, as for KOI 420.01, \citet{borucki11b} found an uncertainty of 1 on the impact parameter, and {\citet{batalha12} report an impact parameter of 0.80$\pm$0.34}, so our observations and models provide a much stronger constraint on the impact parameter, which we find has an associated uncertainty of  0.0771.  Based on our measured planet-star radius ratios and assuming a stellar radius of 0.92 $R_{\odot}$ (from \citealp{batalha12}, and slightly larger than the value of 0.80 $R_{\odot}$ reported by \citealp{borucki11b}), we calculate that KOI 526.01 has a radius between $\sim$ 6.37 and 7.16 $R_{\oplus}$.  {We note that this is over twice the radius of 3.11 $R_{\oplus}$ measured by \citet{batalha12}.}

\subsection{KOI 1187.01}
\label{koi1187res}

As in the case of KOI 225.01, the light curves for KOI 1187.01 (shown in Figure \ref{lc1187}) appear to be fairly V-shaped.  Furthermore, visual inspection of the GTC light curves shows that there is clearly a significant difference in the GTC transit depths.  However, we find that the best-fit planet-star radius ratios only differ at a level of $\sim$1.8$\sigma$, likely due to the large uncertainty in the radius ratio measured for the 858 nm light curve.  We attribute this large uncertainty to a combination of the relatively poor photometric precisions achieved for this target combined with the degeneracy between the impact parameter (measured to be nearly equal to 1) and the planet-star radius ratio.  To reconcile these measurements with what we find visually, we compute weighted mean colors and their uncertainties for the in-transit and out-of-transit data.  We find a mean in-transit color of -0.00425$\pm$0.00050 and a mean out-of-transit color of -0.000160$\pm$0.000212, which differ significantly at a level of $\sim$5.8$\sigma$.  Therefore, despite the consistent measured radius ratios, there is an obviously significant color change during the transit event, so we argue that KOI 1187.01 is in fact a false positive and not a planet.  

Contrary to KOI 225.01, the transit of KOI 1187.01 is deeper in the redder GTC bandpass than in the bluer GTC bandpass.  This implies that during the eclipse, more red light is blocked than blue light, and the smaller (eclipsing) component is bluer than the primary star.  In this case, KOI 1187 may consist of an evolved giant star that is redder and several magnitudes brighter than the eclipsing star.  \citet{slawson11} also list this target in the eclipsing binary catalog as a rejected KOI (which we were also not aware of prior to observing this target).  Also, just as for KOI 225.01, \citet{slawson11} found an orbital period for KOI 1187.01 that is twice as long as that found by \citet{borucki11b}.  As discussed in \S\ref{koi225res} above and below in \S\ref{discuss}, such results have ramifications on any correlation between the orbital period and false positive rate.

\section{Discussion}
\label{discuss}

Of the four $Kepler$ planet candidates presented in this paper, we identified two as false positives and provide further evidence supporting the planet hypothesis for two candidates.  This suggests a false positive rate that is much higher than has been previously predicted for the Borucki et al. (2011b) KOI catalog (Morton \& Johnson 2011; see \S\ref{discuss1} for further discussion), so we consider what could cause such a large false positive rate for our sample.  Referring back to Figure \ref{perhist}, we see that our two false positives, KOI 225.01 and KOI 1187.01, have shorter orbital periods than the two KOIs we validated as planets.  Therefore, our findings suggest that the false positive rate for the $Kepler$ sample varies significantly with orbital period and is largest at the shortest periods ($P$ $<$ 3 days), which is what one would expect $a$ $priori$ due to the rise in the number of detached eclipsing binaries at these short periods (Figure \ref{perhist}).  

We find no significant correlation between the false positive rate and planet radius or other properties of the host star.  Figure \ref{radper} illustrates that there seems to be no trend in the false positive rate with planet radius, as the KOIs we identified as false positives have the smallest and largest apparent radii (as measured by $Kepler$) in our sample.  Similarly, there appears to be no trend with the effective temperature of the host star, although we note that both false positives were also the faintest targets in our sample, as illustrated in Figure \ref{kptemp} (also see \S\ref{discuss1}).  Finally, we emphasize that based on the stellar parameters from the KIC, all our targets are likely FGK dwarfs, which are the primary targets of the $Kepler$ mission. 

One factor that must be considered when making such conclusions is that we do not take into account the level of uncertainty in the stellar parameters that are drawn from the KIC.  Any uncertainty in the stellar parameters would obviously affect the distribution of the planetary radii and stellar magnitudes and temperatures discussed here.  However, if the stellar radii in the KIC have been systematically over- or under-estimated, then all planetary radii would simply scale up or down, and our conclusions about the lack of a correlation between the false positive rate and planet radius would remain the same.\footnote{Some high signal-to-noise ratio eclipsing binaries are more likely to have correctly determined radii, so some short-period planet candidates that are actually eclipsing binaries are correspondingly more likely to have accurate radii.  }

Additionally, there is the issue that if a given planet candidate is actually an eclipsing binary with an orbital period that is twice as long as was initially expected, then this would imply that we are not probing planet candidates with orbital periods of less than 6 days, but instead we are actually probing a sample of eclipsing binaries with periods as long as 12 days.  However, since there is still a comparable population of eclipsing binaries and planet candidates with longer periods (e.g. out to at least 10 days, as illustrated in Figure \ref{perhist}), we believe this period discrepancy would not affect our conclusion that a population of short-period eclipsing binaries can significantly contaminate short-period planet candidates.  Candidates that are initially identified as planets but are actually eclipsing binaries will tend to have shorter apparent periods (their periods will double when they are identified as eclipsing binaries), which strengthens our argument for contamination of the planet sample at short periods, since this effect would further reduce the ratio of planets to eclipsing binaries at periods of less than approximately 3 days.  

Finally, we note that with the technique presented here, it is not possible for us to identify cases where a planet candidate is actually a binary star composed of two stars that have the same temperature, as there would be no measurable color change during the transit event.  This suggests that some of our potentially validated planets could still be false positives.  However, $Kepler$'s photometry should be able to distinguish if the transit depths differ between every other transit (i.e. odd-even transit depths), which would identify a candidate as a false positive.  In cases where a significant odd-even ratio is measured, this can also be used to reconcile the issue with the orbital periods described above.  To the best of our knowledge, no significant odd-even ratio was found for our validated planets, KOI 420.01 and KOI 526.01.

Recently, a new list of $Kepler$ planet candidates was announced, bringing the total number of candidates to 2,321 \citep{batalha12}.  As illustrated in Figures \ref{perhist}$-$\ref{kptemp} and \ref{glat}, the new catalog follows the same general distribution of orbital periods, planet radii, stellar magnitudes, temperatures, and Galactic latitudes as the \citet{borucki11b} catalog, but there is notably a greater number of smaller and shorter-period planet candidates in the new catalog.  This serves to emphasize the need to observationally constrain the false positive rate for such small, short-period planets.  We note that this new catalog was generated using improved vetting metrics, so it should have a higher fidelity than previous ones (i.e., the Borucki et al. 2011b catalog).  However, the KOIs from the \citet{borucki11b} catalog have not been vetted against the improved metrics described in \citet{batalha12}, so both KOI 225.01 and KOI 1187.01 are still included in the updated catalog as potential planet candidates.  Also, \citet{batalha12} remark that potential KOIs are not vetted against the shape of their light curves, so planet candidates with V-shaped transit light curves (such as KOI 225.01 and KOI 1187.01) are not immediately rejected.  Given ongoing parallel efforts to identify planet candidates and KOIs and that \citet{slawson11} previously identified both KOI 225.01 and KOI 1187.01 as eclipsing binaries, we recommend that the eclipsing binary catalog be consulted before following up any given planet candidate.        

\subsection{Comparison to Theoretical Studies}
\label{discuss1}
A recent study by \citet{morton11} provided theoretical estimates of the false positive rate for $Kepler$ planet candidates.  \citet{morton11} specifically estimated that nearly 90\% of the 1,235 candidates presented by \citet{borucki11b} had a false positive probability of less than 10\%.  For our targets in particular, \citet{morton11} computed the following false positive probabilities: 0.01 (KOI 225.01), 0.04 (KOI 420.01), 0.03 (KOI 526.01) and 0.03 (KOI 1187.01).  {It is interesting to note that the target with the lowest false positive probability is one that ended up being a false positive.  However, we note that the false positive probabilities from \citet{morton11} are only valid for non V-shaped transit signals.  Therefore, considering that both KOI 225.01 and KOI 1187.01 appear to have somewhat obviously V-shaped transits, the low false positive probabilities computed for these targets by \citet{morton11} are not appropriate.  A recent study by \citet{morton12} improves upon the analysis from \citet{morton11} by taking the transit shape into account when computing the false positive probability for a given target.  From their new analysis, \citet{morton12} calculate a false positive probability of $>$ 0.99 for KOI 225.01 and 0.76 for KOI 1187.01, which is consistent with our observations.  The caveat about the transit shape also affects the overall assumed false positive rate for the $Kepler$ sample, as \citet{morton11} did not separate planet candidates with V-shaped transit signals from those with non-V-shaped transit signals.  This means that we should not  interpret their findings as 90-95\% of $Kepler$ planet candidates are planets.  However, we can use their averaged false positive probabilities as a starting point for the overall false positive rate for the $Kepler$ sample.  Despite any caveats, we estimate the probability of detecting 2 or 3 false positives out of 4 or 5 targets observed.  Assuming a binomial distribution and a false positive rate of 10\% for $Kepler$ planet candidates \citep{morton11}, we estimate that the probability of detecting 2 false positives from a sample of 4 targets is less than 1\% (and less than 5\% for detecting 3 false positives from 5 targets).  Thus, we use this to emphasize that the false positive rate for current $Kepler$ planet candidates with radii less than $\sim$ 5 $R_{\oplus}$ and orbital periods less than $\sim$ 6.0 days is likely much higher than 10\%.}  

In regards to the full $Kepler$ sample, \citet{morton11} found that the false positive probability varied with the depth of the transit as well as the magnitude and Galactic latitude of the target star, but they did not investigate how the probability might vary with orbital period or transit duration (though they claim that such properties would serve to decrease their probability estimates and thus their estimates are upper limits).  Specifically, our results support their finding that the false positive rate increases slightly with magnitude, as our two faintest targets ended up being false positives.  They also find a general increase in the false positive probability with increasing transit depth (though there are some local minima and maxima; e.g., see Figure 7 in Morton \& Johnson 2011) and with decreasing Galactic latitude.  In our case, the target with the largest transit depth (as measured in the $Kepler$ bandpass) was KOI 420.01, which we validated as a planet.  While we find no obvious correlation in our sample between $Kepler$'s measured transit depths and the false positive rate, dilution is an important factor that has to be considered {(e.g., as illustrated by our light curves for KOI 225.01 in Figure \ref{lc225})}, as some of the depths measured by $Kepler$ may be underestimated.  In regards to Galactic latitude, in Figure \ref{glat} we show the cumulative distribution of the Galactic latitudes for the \citet{borucki11b} KOI list, the \citet{batalha12} KOI list and the \citet{slawson11} catalog of eclipsing binaries, which illustrates that the samples have the same general distribution.  Furthermore, our false positives are indeed located at lower latitudes, which supports the argument by \citet{morton11} that there is a slight increase in the false positive rate with decreasing Galactic latitude.  We present a similar figure in Figure \ref{glat2}, where we show the cumulative distribution of the Galactic latitudes for the $Kepler$ eclipsing binaries along with separate distributions for stars that host short-period ($P$ $\le$ 2 days) and long-period ($P$ $>$ 2 days) planet candidates.  We see that the distribution for the stars that host short-period candidates matches the eclipsing binary distributions more so than the sample of long-period candidates.  Therefore, Figure \ref{glat2} further illustrates the likely contamination of short-period planet candidates by eclipsing binaries.  In Figure \ref{eblat}, we illustrate how the number of eclipsing binaries varies with Galactic latitude, and we find that overall the fraction of $Kepler$ targets that are eclipsing binaries is about the same as a function of Galactic latitude.  This further implies a nearly consistent presence of eclipsing binaries that could plague the $Kepler$ planet candidate sample.

\begin{figure}
\includegraphics[width=84mm]{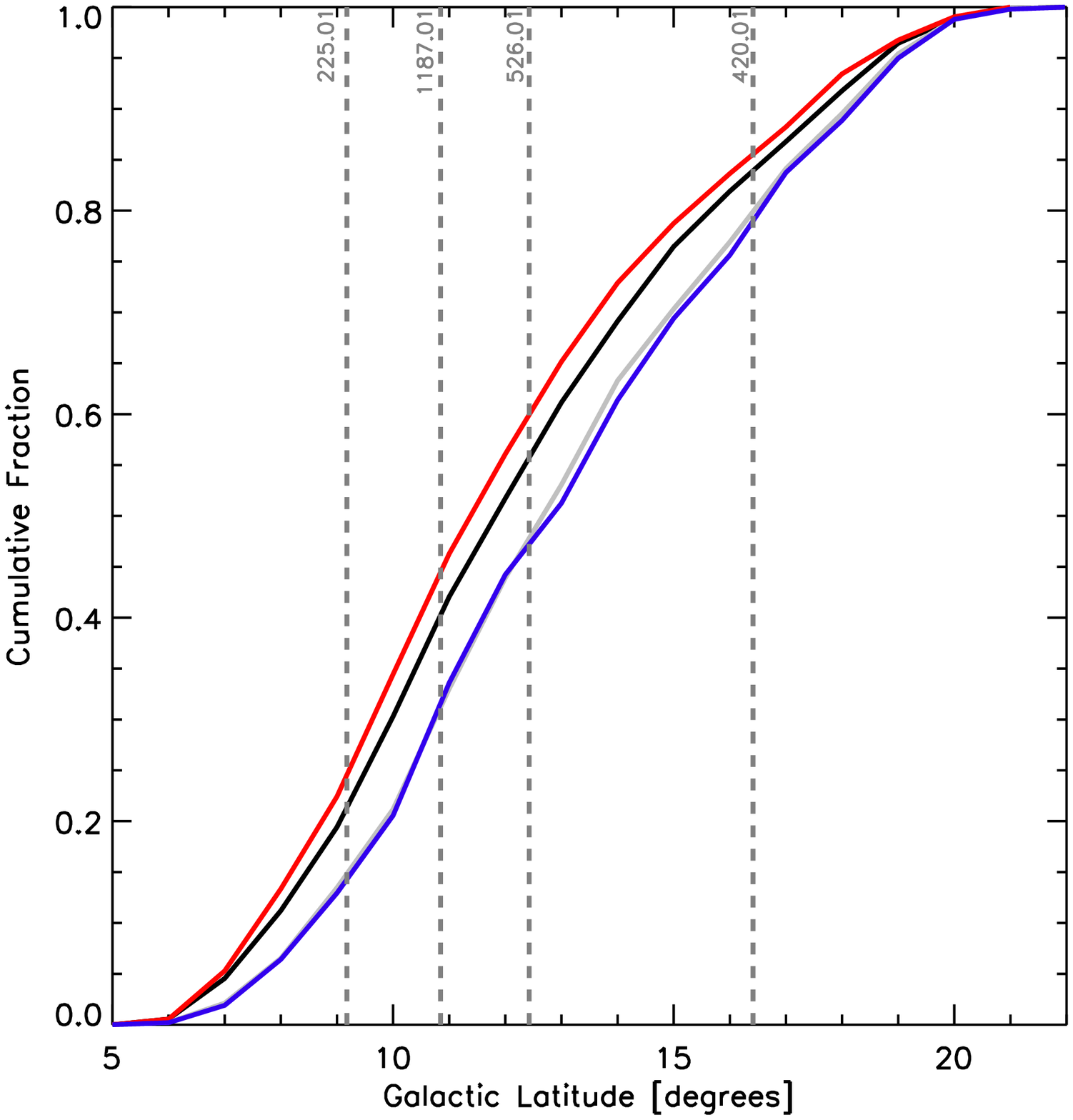}
\caption{Cumulative fraction of planet-hosting stars from \citet[][blue]{borucki11b} and \citet[][gray]{batalha12} and detached (red) and all ``other'' (primarily semi-detached and overcontact; black) eclipsing binaries from \citet{slawson11} as a function of Galactic latitude.  The cumulative functions were normalized against 997 stars for the \citet{borucki11b} sample, 1,790 stars for the \citet{batalha12} sample (which includes the Borucki et al. 2011b sample), 1,266 detached eclipsing binaries, and 901 ``other'' eclipsing binaries (from Slawson et al. 2011).  The gray vertical dashed lines indicate the Galactic latitude of the targets we observed with the GTC, with each target marked accordingly.  The rate of planet-hosting stars is slightly larger towards higher latitudes, while the rate of eclipsing binaries is slightly larger at lower latitudes, but in general the two samples are not significantly different.  Both the KOIs we found to be false positives are located at lower latitudes.}
\label{glat}
\end{figure}

\begin{figure}
\includegraphics[width=84mm]{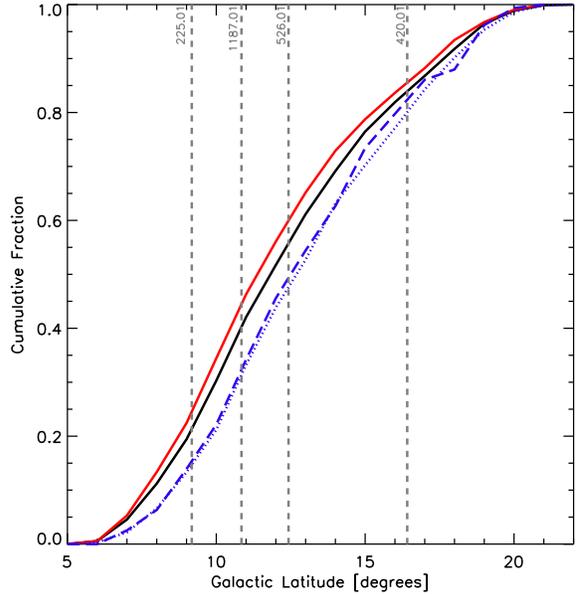}
\caption{Similar to Figure \ref{glat}.  Here, we show the cumulative fraction of stars hosting short-period (P $\le$ 2 days; long-dashed blue curve) and long-period (P $>$ 2 days; dotted blue curve) KOIs (based on the full list of 2,321 KOIs from Batalha et. al 2012) along with the cumulative fraction of two eclipsing binary populations from \citet{slawson11} shown in Figure \ref{glat} as a function of Galactic latitude.  The curves for the planet-hosting stars were normalized against stars hosting 158 short-period and 2,163 long-period planet candidates.  We find that the short-period planet population is more consistent with the eclipsing binary populations than the long-period planet population, which further supports our argument for contamination of the planet population by eclipsing binaries at short periods.  Conversely, the long-period planet population is less likely to be infiltrated by eclipsing binaries.  See the caption to Figure \ref{glat} for additional details.}
\label{glat2}
\end{figure}

\begin{figure}
\includegraphics[width=84mm]{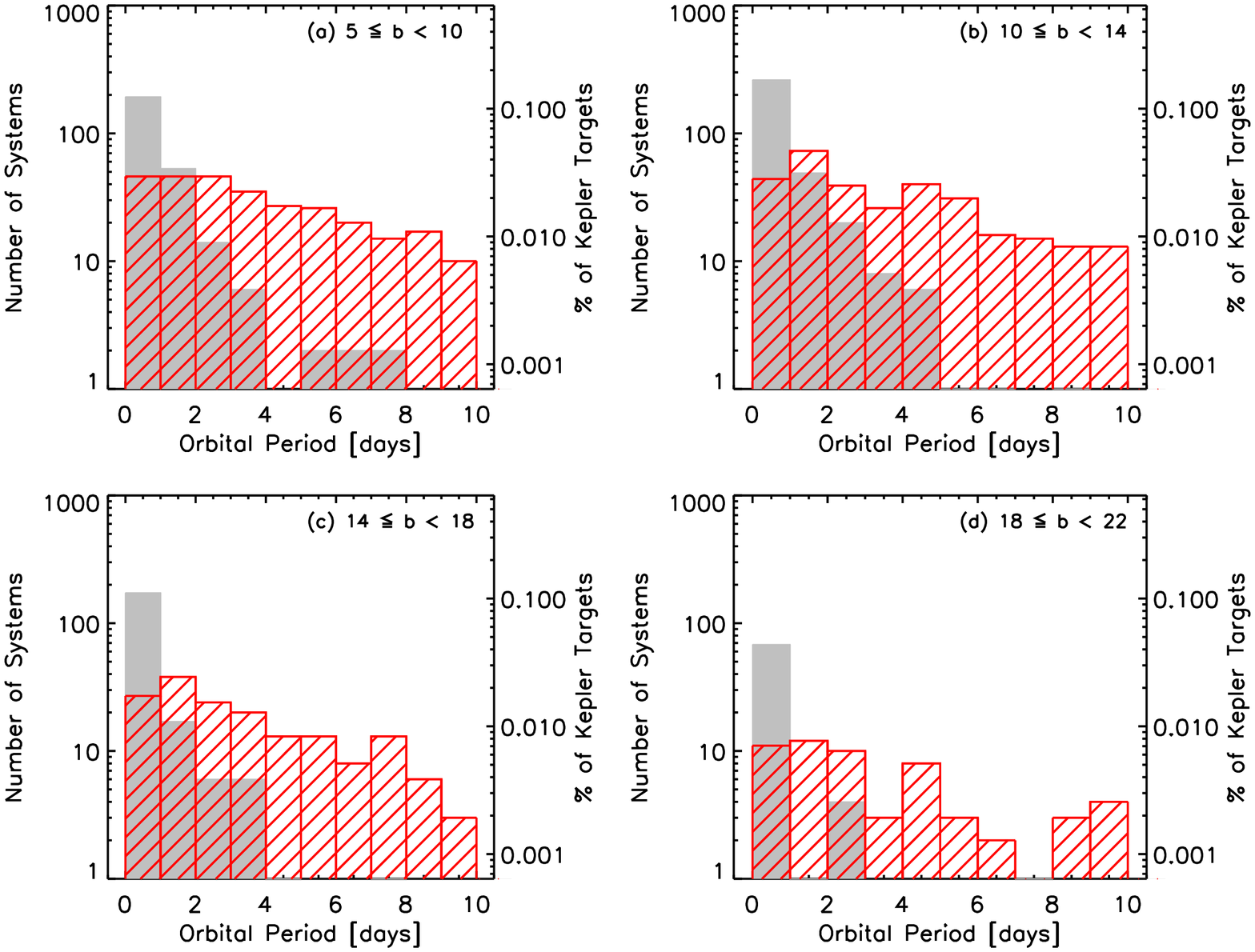}
\caption{Histograms of the number of detached (red) and all ``other'' (gray) eclipsing binaries from \citet{slawson11} as a function of orbital period.  Each panel shows a different bin of Galactic latitude.  As in Figure \ref{perhist}, the righthand y-axis shows the percentage of systems relative to the total number of target stars observed by $Kepler$ in Q1.  There is a roughly consistent number of eclipsing binaries in each latitude bin (except for the highest latitudes), which suggests a consistent presence of eclipsing binaries that could contaminate potential planet candidates.  While there are fewer eclipsing binaries at the highest latitudes, there are also correspondingly fewer planet candidates at those latitudes (see Figure \ref{glat}).}
\label{eblat}
\end{figure}

\subsection{Comparison to Observational Studies}
\label{discuss2}
\citet{desert12} have used warm-$Spitzer$ follow-up of KOIs to find a low false positive rate consistent with the estimates from \citet{morton11}.  Overall, our sample includes shorter period planets than their sample, as well as fainter targets (excluding their sample of M dwarf stars).  Given that only planet candidates with the shortest orbital periods ($<<$ 2 days) are likely to be significantly contaminated by non-detached eclipsing binaries, it is not entirely surprising that \citet{desert12} found a low false positive rate for their sample of planet candidates with periods of $\sim$1.8$-$100 days.

{Recently, \citet{santerne12} presented a Doppler study of short-period giant $Kepler$ planet candidates.  They targeted a sample of 46 $Kepler$ planet candidates with transit depths greater than 0.4\%, orbital periods less than 25 days, and host stars brighter than 14.7 magnitude (Kp).  Based on their radial velocity follow-up observations, \citet{santerne12} estimate a false positive rate of 34.8$\pm$6.5\% for this subset of $Kepler$ planet candidates.  While we probed a different (and smaller) population of $Kepler$ planet candidates, our observed false positive rate of 50\% (with large uncertainty due to our small sample size) for small, short-period $Kepler$ planet candidates is supported by their observations.  The study by \citet{santerne12} also supports our argument that different populations of $Kepler$ targets likely have different false positive rates associated with them.}

\section{Conclusion}
\label{conc}

We acquired multi-color transit photometry of four small ($\sim$2.5$-$4.9 R$_{\oplus}$), short-period ($\sim$0.37$-$6.0 days) $Kepler$ planet candidates with the GTC.  Based on the transit color, we identified two candidates as false positives.  For two, we find further evidence supporting the planet hypothesis, consistent with {\it validated} planets.  We also remind the reader of KOI 565.01, which was a planet candidate observed in a similar fashion and that was also found to be a false positive (albeit it was selected from the first KOI catalog published by Borucki et al. 2011a; Col{\'o}n \& Ford 2011).  While we find a high false positive rate (2/4 or 3/5, if we include KOI 565) in our small sample, we caution that this is likely not representative of the entire sample of $Kepler$ planet candidates, due to the small number of targets we observed and the specific properties of these candidates (e.g. the orbital period and size).  Nevertheless, our results demonstrate the importance of considering these properties when evaluating the false positive probability of specific systems.  While our findings seem to contradict the theoretical estimates from \citet{morton11}, the low false positive rate that they estimate is based on the assumption that all candidates had passed preliminary false-positive vetting metrics based on $Kepler$ photometry and astrometry.  Thus, if we consider the obviously V-shaped transits for KOI 225.01 and KOI 1187.01 to imply a false positive nature for these KOIs, then according to \citet{morton11} the low false positive probabilities for these targets are not accurate.  The false positive rate for our sample is also much larger than the observational constraints from \citet{desert12} that predict that the false positive rate is much less than 10\%.  This is likely partly a result of different targets being probed by the different studies.  {The recent study by \citet{santerne12} further supports this idea, as they found a $\sim$35\% false positive rate for short-period giant $Kepler$ planet candidates}.  We plan to continue observing small, short-period KOIs with the GTC in order to improve the sample size of our study.  The observations presented here, as well as future observations with the GTC, greatly complement follow-up of KOIs done with warm-$Spitzer$ as well as other observatories.  As we can expect the number of $Kepler$ planet candidates to continue to increase, we can use results from all such studies to pinpoint which targets are the best to follow-up in order to maximize the science output from the $Kepler$ mission.

\section*{Acknowledgments}

We gratefully acknowledge the observing staff at the GTC and give a special thanks to Ren\'e Rutten and Antonio Cabrera Lavers for helping us plan and conduct these observations successfully.  We also thank Jean-Michel D\'esert for his feedback throughout this project.  This material is based upon work supported by the National Science Foundation Graduate Research Fellowship under Grant No. DGE-0802270.  This work was also aided by the American Philosophical Society's Lewis and Clark Fund for Exploration and Field Research in Astrobiology and the National Geographic Society's Young Explorers Grant.  This work is based on observations made with the Gran Telescopio Canarias (GTC), installed in the Spanish Observatorio del Roque de los Muchachos of the Instituto de Astrof\'isica de Canarias, on the island of La Palma.  The GTC is a joint initiative of Spain (led by the Instituto de Astrof\'isica de Canarias), the University of Florida and Mexico, including the Instituto de Astronom\'ia de la Universidad Nacional Aut\'onoma de M\'exico (IA-UNAM) and Instituto Nacional de Astrof\'isica, \'Optica y Electr\'onica (INAOE).

\label{lastpage}

\end{document}